\documentclass[prd,aps,showpacs,nofootinbib,tightenlines]{revtex4-1}
\usepackage{amsmath}
\usepackage{bm}
\usepackage{times}
\usepackage{braket}
\usepackage{color}
\usepackage{epsfig}
\usepackage{slashed}
\usepackage{hyperref}
\usepackage{amssymb}
\usepackage{graphicx}
\newcommand{\beq}{\begin{eqnarray}}
\newcommand{\eeq}{\end{eqnarray}}
\newcommand{\non}{\nonumber\\ }
\newcommand{\etab}{\bar{\eta} }

\def \cpc{ Chin. Phys. C  }

\def \epjc{ Eur. Phys. J. C }

\def \jpg{  J. Phys. G }
\def \npb{  Nucl. Phys. B }
\def \plb{  Phys. Lett. B }

\def \prc{  Phys. Rev. C }
\def \prd{  Phys. Rev. D }
\def \prl{  Phys. Rev. Lett.  }

\newcommand{\psl}{ P \hspace{-2.2truemm}/ }

\definecolor{Red}{rgb}{1.,0.,0.}

\definecolor{Blue}{rgb}{0.,0.,1.}

\definecolor{nicered}{rgb}{0.7,0.1,0.1}
\definecolor{nicegreen}{rgb}{0.1,0.5,0.1}

\hypersetup{colorlinks,citecolor=nicegreen,linkcolor=nicered}
\begin{document}

\title{Quasi-two-body decays $B_c \to  D_{(s)} [\rho(770),\rho(1450),\rho(1700) \to ] \pi \pi$ in the perturbative QCD factorization approach}
\author{Ai-Jun Ma$^1$}  \email{theoma@163.com}
\author{Ya Li$^1$}\email{liyakelly@163.com}
\author{Zhen-Jun Xiao$^{1,2}$}\email{xiaozhenjun@njnu.edu.cn}
%
%-----------
\affiliation{$^1$ Department of Physics and Institute of Theoretical Physics,
                          Nanjing Normal University, Nanjing, Jiangsu 210023, P.R. China}
\affiliation{$^2$ Jiangsu Key Laboratory for Numerical Simulation of Large Scale Complex
Systems, Nanjing Normal University, Nanjing, Jiangsu 210023, P.R. China}
\date{\today}
%-----------------------------------------------------%
\begin{abstract}
In this paper, we studied the quasi-two-body
$B_c \to  D_{(s)} [ \rho(770),\rho(1450),\rho(1700) \to ] \pi \pi$ decays
by employing the perturbative QCD (PQCD) factorization approach. The two-pion distribution amplitudes
$\Phi_{\pi\pi}$ are applied to include the final-state interactions between the pion pair,
while the time-like form factors $F_{\pi}(w^2)$ associated with the $P$-wave resonant states $\rho(770)$, $\rho(1450)$
and $\rho(1700)$ are extracted from the experimental data of the $e^+e^-$ annihilation.
We found that:  (a) the PQCD predictions for the branching ratios of the quasi-two-body
$B_c \to  D_{(s)} [\rho(770), \rho(1450), \rho(1700) \to ] \pi \pi$ decays are in the order
of $10^{-9}$ to $ 10^{-5}$ and the direct $CP$ violations around $(10-40)\%$ in magnitude;
(b) the two sets of the large hierarchy $R_{1a,1b,1c}$ and $R_{2a,2b,2c}$ for the ratios of the branching ratios
of the considered decays are defined and can be understood in the PQCD factorization approach, while
the self-consistency between the quasi-two-body and two-body framework for
$B_{c} \to  D_{(s)} [\rho(770)  \to ] \pi \pi$ and $B_{c} \to  D_{(s)} \rho(770)$ decays are confirmed
by our numerical  results;
(c) taking currently known ${\cal B}(\rho(1450)\to\pi\pi)$ and ${\cal B}(\rho(1700) \to \pi\pi)$
as input, we extracted the theoretical predictions for ${\cal B}(B_{c} \to  D \rho(1450)) $
and ${\cal B}(B_{c} \to  D \rho(1700))$ from the PQCD predictions for the decay rates of
the quasi-two-body decays $B_{c} \to  D [\rho(1450), \rho(1700) \to ] \pi \pi$.
All the PQCD predictions will be tested in the future experiments.
\end{abstract}


\maketitle
%------------------------------------------------------%

\section{Introduction}\label{sec:1}

In recent years, large amount of the three-body $B_{(s)}$ decays have been measured~\cite{1612.07233,PDG2016}
and the large localized $CP$ asymmetries in several decay channels~\cite{prl111-101801,prl112-011801,prd90-112004},
have raised great interests. A number of  works have been done by using rather different methods, for example,
the QCD factorization (QCDF)  approach
~\cite{plb622-207,prd74-114009,prd79-094005,prd81-094033,prd88-114014,prd89-074025,prd94-094015,prd89-094007,
npb899-247,epjc75-536,prd87-076007},
the perturbative QCD (PQCD) factorization approach~
\cite{plb561-258,prd70-054006,prd89-074031,plb763-29,prd91-094024,epjc76-675,npb-923,prd95-056008,1701.01844,
1701.02941,1704.07566,1708,1708.02869},
and the frameworks based on the symmetry principles ~\cite{prd72-075013,prd72-094031,prd84-056002,
plb726-337,prd89-074043,plb728-579,prd91-014029,prd84-034040}.
 Compared to $B_{(s)}$ meson, $B_c$ meson is unique since it consists of two different heavy quarks: $\bar{b}$
 and $c$ quark. With the flavor quantum
numbers $B =-C = \pm1$, the $B_c$ meson can not decay strongly but only weakly. Besides, it is heavier than
$B_{(s)}$ meson and more difficult to be produced unless in high energy hadron collisions. Fortunately, some
$B_c$ events have been observed in the Tevatron and Large Hadron Collider (LHC) experiments~\cite{1612.07233,PDG2016}.
In recent works by the LHCb Collaboration, some three-body $B_c$ decays, for instance, $B_c^+ \to
 \{KKK, \pi\pi\pi, KK\pi, p\bar{p}K, p\bar{p}\pi\}$ have been measured~\cite{prd94-091102,plb759-313}.
 Meanwhile, more and more $B_c$ events will be collected with the continuous running of LHC. The research
 of the three-body $B_c$ decays could be an important topic for both experiment and theory in next few years.

As known, in three-body decays, one can measure the distribution of $CP$ asymmetry in the Dalitz plot~\cite{pm44-1068} experimentally.
However, from a theoretical point of view, it is too difficult to calculate $CP$ violation in the whole Dalitz
plot but practical to
analyze a process of quasi-two-body decay.
Experimentally, three-body $B$ meson decays are known to be usually dominated by the low energy resonances on
$\pi\pi$, $KK$ and $K\pi$ channels
 and most of the quasi-two-body decays are extracted from the Dalitz-plot analysis of three-body ones.
In a quasi-two-body decay, the final-state interactions between the pair of mesons are considered while the
rescattering between the third particle and the meson pair is usually ignored. In the views of
PQCD~\cite{plb561-258,prd70-054006}, a direct evaluation of the hard kernels
which contain two virtual gluons at lowest order is not important, the dominant contributions
come from the region where the
two energetic light mesons are almost collimating to each other with an invariant mass below
$O(\bar\Lambda m_B)$($\bar\Lambda=m_B-m_b$), and
the two-meson distribution amplitudes~\cite{plb561-258,prd70-054006,Muller,Grozin,prl81-1782,npb555-231}
have been introduced to include both resonant and nonresonant contributions for the meson pair.
In the previous work, the parameters in the $P$-wave two-pion distribution amplitudes were
determined in PQCD approach~\cite{plb763-29}. Following Ref.~\cite{plb763-29}, we have
studied the quasi-two body decays
$B_{(s)} \to P/D [\rho(770),\rho(1450),\rho(1700)\to] \pi \pi$~\cite{prd95-056008,1704.07566,npb-923,1708}
where $P = \pi, K, \eta ,\eta^\prime$, and $D$ represents the charmed $D$ meson.

In the past several years, a series of semileptonic $B_c$ decays~\cite{prd90-094018} and nonleptonic
two-body $B_c$ decays~\cite{prd81-014022,prd81-074017,prd81-037501,prd81-074010,prd82-054029,JPG38-035009,
prd84-074033,prd86-074008,prd87-074027,prd86-074019,prd96-013005}  have been studied in the PQCD framework.
End-point singularity is avoid by keeping the transverse momentum $k_T$ of the quarks, and the Sudakov
formalism makes this approach more reliable.
From those literatures, we know the following points which can be also helpful for us to study the three-body $B_c$ decays:
\begin{itemize}
\item[(1)]
 The size of annihilation contributions is a meaningful issue
in $B_c$ physics since the two-body nonleptonic charmless decays $B_c \to h_1 h_2$ ($h_1$, $h_2$ represent
the light pseudoscalar mesons, vector mesons, axial-vector mesons,  scalar mesons and so on) occur through
the weak annihilation diagrams only. As a feature of PQCD, the diagrams including factorizable, nonfactorizable
and annihilation type are all calculable. From numerical calculation, the contribution from nonfactorizable and
annihilation-type diagrams is also found to be of great importance in charmed decays $B_c \to D h$
($D$ stands for charmed $D$ meson);
\item[(2)]
Since only tree operators are involved, the direct $CP$-violating asymmetries for
those charmless $B_c$ decays are absent naturally, while there are both penguin and tree diagrams involved in
$B_c \to D h$ decays and the possibly large
direct $CP$ violations in some channels were predicted ~\cite{prd86-074008,prd87-074027}.
\end{itemize}

In this work, we will extend the previous studies as presented in
Refs.~\cite{plb763-29,npb-923,prd95-056008,1704.07566,1708} to the quasi-two-body decays
$B_c \to  D_{(s)} [\rho(770) ,\rho(1450), \rho(1700) \to] \pi \pi$, and give our predictions
about the branching ratios and direct $CP$ violations of those decays. For simplicity, we generally
use the abbreviation $\rho=\rho(770)$, $\rho'=\rho(1450)$, $\rho''=\rho(1700)$ in the following
sections. By now, the $B_c \to  D_{(s)} \rho$ decay has been studied in several frameworks, for
examples, a relativistic constituent quark model based on the
Bethe-Salpeter formalism~\cite{prd56-4133}, the QCD factorization approach with input from light-front
quark model~\cite{prd80-114003} and the PQCD approach~\cite{prd86-074008}. But there are hardly any
studies for $B_c$ decays with  final states $\rho'$ and $\rho''$ since the structure of the excited
states $\rho'$ and $\rho''$ is not yet completely clear~\cite{PDG2016}. There are a small number of
theoretical studies for $\rho'$ and $\rho''$, for examples, the
Refs.~~\cite{prd60-094020,prc79-025201,prd77-116009,1205-6793,prd88-093002}.
Experimentally, the observation of both $\rho'$ and $\rho''$ has been reported in study of the
$\tau^- \to \pi^-\pi^0\nu_\tau$ decay by Belle~\cite{prd78-072006}
and $e^+ e^- \to \pi^+ \pi^-(\gamma)$ decay by {\it BABAR}~\cite{prd86-032013}. Meanwhile,
several quasi-two-body $B$ meson decays like $B^0 \to K^+ \rho'^- $, $B^- \to \pi^- \rho'^0 $
and $B^0 \to \bar{D^0} \rho'^0 (\rho''^0)$
have been observed in experiments~\cite{prd83-112010,prd79-072006,prd92-032002}. For the
phenomenological analysis of $B_c \to  D_{(s)} \rho' (\rho'')$, we can not treat it as in
Refs~\cite{prd81-014022,prd81-074017,prd81-037501,prd81-074010,prd82-054029,JPG38-035009,
prd84-074033,prd86-074008,prd87-074027,prd86-074019,prd96-013005} due to the lack of the
distribution amplitudes for $\rho'$ and $\rho''$.  But in quasi-two-body framework~\cite{plb763-29},
we here firstly attempt to study the $B_c \to  D_{(s)} [\rho' (\rho'') \to] \pi\pi$ decays by singling
out the component of $\rho'(\rho'')$ in the two-pion distribution amplitudes. Then, the branching
ratios of $B_c \to  D_{(s)} \rho' (\rho'')$ will be extracted from the results of $B_c \to  D_{(s)} [\rho' (\rho'') \to] \pi\pi$
relying on a reliable estimation for the branching fraction ${\cal B}(\rho'(\rho'') \to \pi\pi)$.

The paper is organized as follows. In Sec.~II, we give a brief introduction for the theoretical
framework and perturbative calculations for the considered decays. Then, the numerical values
and phenomenological analysis are given in Sec.~III. Finally, the last section contains a short summary.

\section{The theoretical framework}\label{sec:2}

\begin{figure}[tbp]
\begin{center}
\vspace{-3cm} \centerline{\epsfxsize=17cm \epsffile{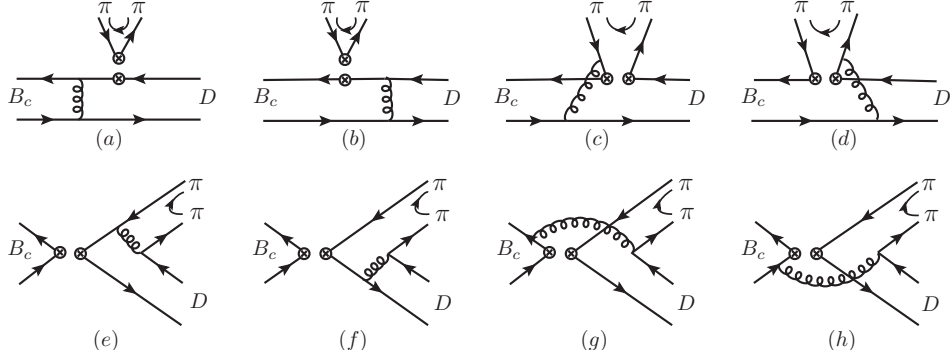}}
\vspace{-17cm} \caption{Typical Feynman diagrams for the $B_c \to  D_{(s)} [\rho ,\rho', \rho'' \to] \pi \pi$ decays.}
\label{fig:fig1}
\end{center}
\end{figure}

In the PQCD approach based on $k_T$ factorization, one separates the hard
and soft dynamics in a QCD process~\cite{plb561-258}. The hard part is calculable in the perturbation
theory while the soft part is not calculable perturbatively but have to be treated as an universal input determined from
experiments. The amplitude of the process, consequently, could be expressed as a convolution of a hard kernel
$H$ with the hadron wave functions $\Phi(x, k_T )$ ($x$ means a longitudinal momentum fraction and $k_T$ represents
a transverse momentum).
For a quasi-two-body $B_c$-meson decay, its decay amplitude $\cal A$ in PQCD approach can then be written
conceptually as the following convolution~\cite{plb561-258,prd70-054006}
\begin{eqnarray}
{\cal A}=\Phi_{B_c}\otimes H\otimes \Phi_{h_1 h_2}\otimes\Phi_{h_3}.
\end{eqnarray}
The symbols $\otimes$ means the convolution integrations over the
parton kinematic variables and the specific calculation formula will be shown in the following subsections.

\subsection{Coordinates and wave functions}\label{sec:21}

In the light-cone coordinates, the $B_c$ meson momentum $p_{B}$,
the momenta  $p_1$, $p_2$ for each pion and the total momentum of the pion pair $p=p_1+p_2$,
and the $D$ meson momentum $p_3$ in the rest frame of $B_c$ meson are chosen as
\begin{eqnarray}\label{lc}
p_{B}&=&\frac{m_{B_c}}{\sqrt2}(1,1,\textbf{0}_{\rm T}),
~\quad p=\frac{m_{B_c}}{\sqrt2}(1-r^2,\eta,\textbf{0}_{\rm T}),~\quad
p_3=\frac{m_{B_c}}{\sqrt2}(r^2,1-\eta,\textbf{0}_{\rm T}), \non
p_1&=& \frac{m_{B_c}}{\sqrt2} \left ( \zeta (1-r^2), (1-\zeta)\eta, \textbf{0}_{\rm T}\right), \quad
p_2= \frac{m_{B_c}}{\sqrt2} \left( (1-\zeta)(1-r^2), \zeta\eta, \textbf{0}_{\rm T}\right),
\end{eqnarray}
where $\eta=w^2/[(1-r^2)m^2_{B_c}]$ with the mass ratio $r=m_{D}/m_{B_c}$ and the invariant mass squared of the
pion pair $w^2=p^2=m^2(\pi\pi)$,  $\zeta$ is the momentum fraction for one of the pion pair.
The momenta of the light quarks in the $B_c$ meson and the final state mesons are defined as
$k_B$, $k$ and $k_3$ respectively
\begin{eqnarray}
k_B=x_Bp_B+(0,0,\textbf{k}_{B \rm T}),~\quad k=zp^++(0,0,\textbf{k}_{\rm T}),~\quad
k_3=x_3p_3^-+(0,0,\textbf{k}_{3\rm T}),
\end{eqnarray}
where the momentum fraction $x_{B}$, $z$ and $x_3$ run between zero and unity.

For $B_{c}$ meson, we use the same wave function as in
Refs.~\cite{prd81-014022,prd81-074017,prd81-037501,prd81-074010,prd82-054029,
JPG38-035009,prd84-074033,prd86-074008,prd87-074027,prd86-074019,prd96-013005}:
\begin{eqnarray}
\Phi_{B_{c}}(x)=\frac{i}{\sqrt{6}}(\psl_B \; + m_{B_{c}})\gamma_{5} \phi_{B_c}(x,b),
\end{eqnarray}
with the distribution amplitude $\phi_{B_c}(x,b)$ \cite{prd86-074019}
\begin{eqnarray}
\phi_{B_c}(x,b)=\frac{f_{B_c}}{2\sqrt{6}}\,\delta \left (x- \frac{m_c}{m_{B_c}} \right)
\exp\left[-\frac{1}{2}\omega^2_{B_c}b^2\right],
\end{eqnarray}
where the exponent term describes the $k_{\rm T}$-dependence of $\phi_{B_c}(x,b)$;
while the parameter $\omega_B=(0.60 \pm 0.05)$ GeV, $m_c$ is the charm quark mass,
$m_{B_c}$ is the $B_c$ meson mass, and $f_{B_c}$ is the decay constant of $B_c$ meson.

For $D$ meson, the two-parton light-cone distribution amplitudes in the heavy quark limit
can be written as \cite{prd86-074019,prd86-074008,prd87-074027,prd78-014018,prd86-094001,prd95-016011,epjc45-711}
\begin{eqnarray}
\langle D(p_3)|q_{\alpha}(z)\bar{c}_{\beta}(0)|0\rangle
\,&=&\,\frac{i}{\sqrt{6}}\int_{0}^{1}dx\,e^{ixp_3\cdot
z}\left[\gamma_{5}(\makebox[-1.5pt][l]{/}p\,_3+\,m_{D})\phi_{D}(x,b)\right]_{\alpha\beta},
\end{eqnarray}
with the distribution amplitude  $\phi_{D}(x,b)$
\begin{eqnarray}
\phi_{D}(x,b)=\frac{1}{2\sqrt{6}}\,f_{D}\,6x(1-x)\left[ 1+C_{D}(1-2x) \right]
\exp\left[\frac{-\omega^{2}b^{2}}{2}\right],
\label{eq:phidxb}
\end{eqnarray}
where $C_{D}=0.5\pm0.1, \omega=0.1$ GeV and $f_{D}=211.9$ MeV~\cite{PDG2016} for the $D$ meson,
and $C_{D_s}=0.4\pm0.1, \omega=0.2$ GeV and $f_{D_{s}}=249$ MeV~\cite{PDG2016} for $D_{s}$ meson .

The $P$ wave two-pion distribution amplitudes are defined in the same way as in Ref.~\cite{plb763-29}:
\begin{eqnarray}
\Phi^{I=1}_{\pi\pi}=\frac{1}{\sqrt{2N_c}}\left[p \hspace{-2.0truemm}/\phi^0(z,\zeta,w^2)
+w\phi^s(z,\zeta,w^2) +\frac{p \hspace{-2.0truemm}/_ 1p \hspace{-2.0truemm}/_2
-p \hspace{-2.0truemm}/_2p \hspace{-2.0truemm}/_ 1}{w(2\zeta-1)} \phi^t(z,\zeta,w^2)\right]\; ,
\end{eqnarray}
where
\begin{eqnarray}
\phi^0(z,\zeta,w^2)&=& \frac{3F_\pi(w^2)}{\sqrt{2N_c}}z(1-z)\left [ 1+a^0_2 C_2^{3/2}(t)\right ]P_1(2\zeta-1),\nonumber\\
\phi^s(z,\zeta,w^2)&=& \frac{3F_s(w^2)}{2\sqrt{2N_c}}(1-2z)\left [1+a^s_2 (1-10z+10z^2) \right ]P_1(2\zeta-1),\nonumber\\
\phi^t(z,\zeta,w^2)&=& \frac{3F_t(w^2)}{2\sqrt{2N_c}}(1-2z)^2\left [1+a^t_2 C_2^{3/2}(t) \right ]P_1(2\zeta-1),
\end{eqnarray}
with $C_2^{3/2}(t)=\frac{3}{2}(5t^2-1)$, $t=1-2z$ and
the Legendre polynomial $P_1(2\zeta-1)=2\zeta-1$. The Gegenbauer moments are chosen as $a^0_2=0.30\pm0.05$, $a^s_2=0.70\pm0.20$
and $a^t_2=-0.40\pm0.10$~\cite{prd95-056008}. The time-like form factor
$F_\pi$ which includes the strong interactions between the $P$-wave resonances
and the pion pair can be written in the form of~\cite{prd86-032013}
\begin{eqnarray}
F_\pi(w^2) &=& \frac{1}{1+\sum_i c_i}\cdot \left \{ {\rm BW}^{\rm GS}_\rho (w^2, m_\rho, \Gamma_\rho)
\frac{1+c_\omega {\rm BW}^{\rm KS}_\omega(w^2, m_\omega, \Gamma_\omega)}{1+c_\omega}
+ \sum_i c_i   {\rm BW}^{\rm GS}_i   (w^2, m_i,  \Gamma_i) \right\}, ~~\label{eq:fpiw2}
\end{eqnarray}
where $i=(\rho', \rho'', \rho(2254))$.
The explicit expressions of  ${\rm BW}^{\rm GS}_{\rho,i},{\rm BW}^{\rm KS}_\omega$, and relevant parameters can be also
found for example in Ref.~\cite{prd86-032013}.

\subsection{Decay amplitudes}\label{sec:22}

For the considered $B_c \to  D_{(s)}  [\rho,\rho',\rho'' \to ] \pi \pi$ decays, the effective Hamiltonian
$H_{eff}$~\cite{rmp68-1125} can be written as:
\begin{eqnarray}
\mathcal{H}_{eff}=\frac{G_{F}}{\sqrt{2}} \left\{ \sum_{q=u,c}V_{qb}^{*}V_{qd(s)}
\left[C_{1}(\mu)O_{1} (\mu)+C_{2}(\mu)O_{2}(\mu)\right] -V_{tb}^{*}V_{td(s)}\sum_{i=3}^{10}C_{i}(\mu)O_{i}(\mu)\right\},
\end{eqnarray}
where $G_F=1.16639\times 10^{-5}$ GeV$^{-2}$ is the Fermi coupling constant,
$C_i(\mu)$ are the Wilson coefficients at the renormalization scale
$\mu$, $O_i(\mu)$ are the effective four quark operators and $V_{ij}$ are the CKM matrix elements.

At the leading order, there are eight diagrams contributing to
the considered decays as shown in the Fig.~\ref{fig:fig1}. The four diagrams in first line are the emission
type diagrams while the diagrams in the second line are the four annihilation type diagrams.
By making analytical evaluations for those Feynman diagrams in Fig.~\ref{fig:fig1}, we can obtain
the total decay amplitudes of these considered decays.

For the three $\rho(770)$-related  $B_c \to D_{(s)} [\rho \to ] \pi\pi$ decays, their total decay amplitudes
can be written in the following form
\begin{eqnarray}
{\cal A }({B_c^+ \to D^0 [\rho^+  \to] \pi^+\pi^0})&=&\frac{G_F}{\sqrt2}
\Big \{ V^*_{ub}V_{ud}
\left [ a_1 \; F_{eD}^{LL}+C_1M_{eD}^{LL} \right ]
+V^*_{cb}V_{cd} \left [ a_1 \; F_{aD}^{LL}+C_1M_{aD}^{LL} \right ]\non
&&-V^*_{tb}V_{td} \Big [ ( a_4+ a_{10} )\left ( F_{eD}^{LL}+F_{aD}^{LL} \right )
+(a_6+ a_8 )\left (F_{eD}^{SP} +F_{aD}^{SP} \right )\non
&&+( C_3+C_9)\left ( M_{eD}^{LL}+M_{aD}^{LL} \right )+(C_5+C_7)\left ( M_{eD}^{LR}+M_{aD}^{LR} \right )\Big]
\Big \}, \label{eq:a01}
\end{eqnarray}
\begin{eqnarray}
\sqrt2{\cal A}({B_c^+ \to D^+ [\rho^0  \to] \pi^+\pi^-})&=&
\frac{G_F}{\sqrt2} \Big \{ V^*_{ub}V_{ud} \left [ a_2\; F_{eD}^{LL}+ C_2 M_{eD}^{LL} \right ]
-V^*_{cb}V_{cd} \left [ a_1\; F_{aD}^{LL}+C_1M_{aD}^{LL} \right ]\non
&& -V^*_{tb}V_{td} \Big [ \left (-a_4+ \frac{5}{3}C_9+C_{10} \right ) F_{eD}^{LL}
+ \frac{3}{2}a_7 \; F_{eD}^{LR} -\left( a_6-\frac{1}{2}a_8 \right ) F_{eD}^{SP}\non
&&
+\left ( -C_3+\frac{3}{2}a_{10} \right ) M_{eD}^{LL}
-\left ( C_5-\frac{1}{2}C_7 \right ) M_{eD}^{LR}
+\frac{3}{2}C_8 M_{eD}^{SP}\non
&&  -(a_4+a_{10} )F_{aD}^{LL}-( a_6+a_8 )F_{aD}^{SP}-(C_3+C_9)M_{aD}^{LL}-(C_5+C_7)M_{aD}^{LR} \Big ] \Big \},
\label{eq:a02}
\end{eqnarray}
\begin{eqnarray}
\sqrt2\mathcal{A}({B_c^+ \to D_s^+ [\rho^0 \to] \pi^+\pi^-})&=&
\frac{G_F}{\sqrt2} \Big \{ V^*_{ub}V_{us} \left [ a_2\; F_{eD}^{LL}+C_2M_{eD}^{LL} \right ] \non
&& -V^*_{tb}V_{ts}\frac{3}{2}\Big [ a_9\; F_{eD}^{LL}+ a_7 \; F_{eD}^{LR}+ C_{10} M_{eD}^{LL}+C_8 M_{eD}^{SP} \Big ] \Big\},
\label{eq:a03}
\end{eqnarray}
where $a_i$ are the combinations of the Wilson coefficients $C_i$,
\beq
a_{1,2}&=&C_{2,1}+\frac{C_{1,2}}{3}, \non
a_i&=&C_i+\frac{C_{i\pm 1}}{3}, {\rm for}\ \ i=(3,5,7,9); \ \ {\rm or} \ \ i=(4,6,8,10).
\eeq
The $F_{eD}^{LL}$ \footnote{The subscripts $LL$, $LR$ and $SP$ correspond to the contributions from the $(V-A)(V-A)$,
$(V-A)(V+A)$ and $(S-P)(S+P)$ currents, respectively.} and other individual amplitudes
relevant with the eight sub-diagrams in Fig.~\ref{fig:fig1}  can be written in the following forms:

(1) From the factorizable emission diagrams Fig.~\ref{fig:fig1}(a) and \ref{fig:fig1}(b):
\begin{eqnarray}
 F_{eD}^{LL}&=&8\pi C_F m^4_{B_c} F_\pi \int_0^1 dx_B dx_3
\int_0^{1/\Lambda} b_B db_B \; b_3 db_3\; \phi_B \phi_D\nonumber\\
&& \times \Big\{\Big[-\etab [\eta(1-x_3)+(r-2)r_b+x_3(1-2r)] +r^2(x_3-2r_b)
\Big] E_e(t_a) h_a(\alpha,\beta,b_3,b_B)\; S_t(x_3 \etab )\non
&& +\Big[-\etab [ 2r(x_B-1)+\eta x_B]+r^2(x_B-1)\Big]  E_e(t_b) h_b(\alpha,\beta,b_B,b_3)\; S_t(|r^2-x_B|)\Big\},\\
F_{eD}^{LR}&=&F_{eD}^{LL}, F_{eD}^{SP}=0.
\end{eqnarray}

(2) From the nonfactorizable emission diagrams Fig.~\ref{fig:fig1}(c) and \ref{fig:fig1}(d):
\begin{eqnarray}
M_{eD}^{LL}&=&-32\pi C_F m^4_{B_c}/\sqrt{6} \int_0^1 dx_B dz dx_3
\int_0^{1/\Lambda} b_B db_B\; b db\; \phi_B\phi_D \phi^0\; \non
&& \times \Big\{\Big[r[(1+\eta)(1-x_B)-\etab x_3-\eta z] -(1-\eta^2)(1-x_B-z)\Big] E_n(t_c) h_c(\alpha,\beta,b_B,b)\non
&& + \Big[(\etab-r)(1-x_3)\etab +r[(1+\eta)x_B -\eta z] + \etab (z-2x_B)\Big] E_n(t_d) h_d(\alpha,\beta,b_B,b) \Big\},
\end{eqnarray}
\begin{eqnarray}
M_{eD}^{LR}&=&32\pi C_F m^4_{B_c}/\sqrt{6} \int_0^1  dx_B dz dx_3
\int_0^{1/\Lambda} b_B db_B\; b db\; \phi_B\phi_D \sqrt{\eta(1-r^2)}\nonumber\\
&&\times\Big\{\Big[ \etab (1-x_B-z)(\phi^s+\phi^t)-r( \etab x_3-z)(\phi^s-\phi^t)+2r(1-x_B-z)\phi^s\Big] E_n(t_c)h_c(\alpha,\beta,b_B,b)\non
&& -\Big[ r[ \etab (1-x_3)-z](\phi^s+\phi^t)-\etab (x_B-z)(\phi^s-\phi^t)+2(z-x_B)\phi^s \Big] E_n(t_d)h_d(\alpha,\beta,b_B,b) \Big\},
\end{eqnarray}
\begin{eqnarray}
 M_{eD}^{SP}&=&-32\pi C_F m^4_{B_c}/\sqrt{6} \int_0^1 dx_B dz dx_3
\int_0^{1/\Lambda} b_B db_B\; b db\; \phi_B\phi_D\phi^0\non
&&\times\Big\{\Big[(\etab-r) \etab x_3+r[(1+\eta)(1-x_B)-\eta z]+ \etab [z-2(1-x_B)]\Big]
E_n(t_c)h_c(\alpha,\beta,b_B,b)\non
&& -\Big[r[ \etab (1-x_3)+\eta(z-x_B)-x_B] +(1-\eta^2)(x_B-z)\Big] E_n(t_d)h_d(\alpha,\beta,b_B,b) \Big\}.
\end{eqnarray}

(3) From the factorizable annihilation diagrams Fig.~\ref{fig:fig1}(e) and \ref{fig:fig1}(f):
\begin{eqnarray}
F_{aD}^{LL}&=&-8\pi C_F m^4_{B_c} f_{B_c}\int_0^1 dz dx_3
\int_0^{1/\Lambda} b db\; b_3 db_3\;\phi_D\non
&&\times\Big\{\Big[\etab [(1-r^2) \etab x_3+\eta]\phi^0+ 2r\sqrt{\eta(1-r^2)} \left [1+\eta + \etab x_3 \right ] \phi^s \Big]
E_a(t_e)h_e(\alpha,\beta,b_3,b)S_t(x_3 \etab )\non
&&  +\Big[ \left [2(1+\eta)rr_c- \etab z+ r^2(2 \etab z - 1 ) \right ]\phi^0
-\sqrt{\eta(1-r^2)} \left [2rz(\phi^s+\phi^t)+(2r-r_c) \etab (\phi^s-\phi^t) \right ] \Big]\non
&&\times  E_a(t_f) h_f(\alpha,\beta,b,b_3)S_t(z \etab ) \Big\},
\end{eqnarray}
\begin{eqnarray}
 F_{aD}^{SP}&=&16\pi C_F m^4_{B_c} f_{B_c}\int_0^1 dz dx_3
\int_0^{1/\Lambda} b db\; b_3 db_3\;\phi_D \non
&& \times \Big\{\Big[r (x_3 \etab +2\eta)\phi^0+2\sqrt{\eta(1-r^2)} \etab \phi^s  \Big]
E_a(t_e) h_e(\alpha,\beta,b_3,b) S_t(x_3\etab)\non
&& +\Big[ [2r(1-(1-z)\eta)- \etab r_c]\phi^0+\sqrt{\eta(1-r^2)}
\left [ \etab z(\phi^s-\phi^t) -4rr_c\phi^s \right ] \Big] E_a(t_f)h_f(\alpha,\beta,b,b_3)S_t(z \etab ) \Big\}.
\end{eqnarray}

(4) From the nonfactorizable annihilation diagrams Fig.~\ref{fig:fig1}(g) and \ref{fig:fig1}(h):
\begin{eqnarray}
 M_{aD}^{LL}&=&32\pi C_F m^4_{B_c}/\sqrt{6} \int_0^1  dx_B dz dx_3
\int_0^{1/\Lambda} b_B db_B\; b db\; \phi_B\phi_D\non
&&\times\Big\{\Big[ \etab [(1+\eta)(1-x_B-z)-r_b]\phi^0 +r\sqrt{\eta(1-r^2)} [-z(\phi^s+\phi^t)-  \etab x_3(\phi^s-\phi^t)\non
&& +2(1-x_B-2r_b)\phi^s]\Big]
E_n(t_g) h_g(\alpha,\beta,b,b_B) \non
&& +\Big[( \etab [ \etab x_3 -(1+\eta)x_B+r_c+\eta z]\phi^0+r\sqrt{\eta(1-r^2)}[ \etab x_3(\phi^s+\phi^t)
+z(\phi^s-\phi^t)\non
&& +2(2r_c-x_B)\phi^s]\Big] E_n(t_h) h_h(\alpha,\beta,b,b_B)  \Big\},
\end{eqnarray}
\begin{eqnarray}
 M_{aD}^{LR}&=&-32\pi C_F m^4_{B_c}/\sqrt{6} \int_0^1 dx_B dz dx_3
\int_0^{1/\Lambda} b_B db_B\; b db \;\phi_B\phi_D\non
&&\times\Big\{\Big[[(1+\eta)(1+r_b-x_B)- \etab x_3-\eta z]\phi^0- \etab \sqrt{\eta(1-r^2)}(1+r_b-x_B-z)(\phi^s+\phi^t)\Big]  \non
&&\times E_n(t_g)h_g(\alpha,\beta,b,b_B)\non
&& -\Big[r \left[(1+\eta)(x_B+r_c)- \etab x_3 -\eta z \right ]\phi^0
- \etab \sqrt{\eta(1-r^2)} \ (r_c+x_B-z)(\phi^s+\phi^t)\Big ]\non
&& \times E_n(t_h)h_h(\alpha,\beta,b,b_B) \Big\},\label{eq:a1}
\end{eqnarray}
where $\etab=1-\eta$, $C_F=4/3$ is the color factor. The explicit expressions of the hard functions
$(h_a,\cdots,h_h)$, the hard scales $(t_a,\cdots,t_h)$,  the evolution factors
$(E_a, \cdots,E_n)$ and the threshold resummation factor $S_t(x_i)$ will be given in Appendix \ref{sec:app}.

For the decays involving $\rho'$ and $\rho''$ mesons, one can get the relevant expressions
for the corresponding decay amplitudes by simple replacements of $\phi^{0,s,t}$ for $\rho$ meson to
the ones for $\rho'$ or $\rho''$, respectively.

%%%%%%%%%%%%%%%%%%%%%%%%%%%%%%%%%%%%%%%%%%%%%%%%%%%

\section{Numerical results}\label{sec:3}

Besides those specified in previous sections, the following input parameters will also be used in our numerical
calculations (the masses, decay constants and QCD scale are in units of GeV) ~\cite{PDG2016}:
\begin{eqnarray}
\Lambda^{(f=4)}_{ \overline{MS} }&=&0.25, \quad m_{B_c}=6.275,
\quad m_{D^+}=1.870,\quad m_{D^0}=1.865,\quad m_{D_s^+}=1.968,\nonumber\\
 \quad m_{\pi^\pm}&=&0.140, \quad m_{\pi^0}=0.135, \quad
m_{b}=4.8, \quad f_{B_c}= 0.489, \quad \tau_{B_c}= 0.507\; {\rm ps}. \label{eq:inputs}
\end{eqnarray}
For the Wolfenstein parameters $(A, \lambda,\bar{\rho},\bar{\eta})$ of the CKM mixing matrix, we use
$A=0.811 \pm 0.026,~\lambda=0.22506\pm 0.00050$,~$\bar{\rho} = 0.124_{-0.018}^{+0.019},~\bar{\eta}= 0.356\pm 0.011$.

For the considered $B_c \to  D_{(s)} [\rho, \rho', \rho'' \to ] \pi \pi$ decays, the differential decay rate
can be written in the following form
\begin{eqnarray}
\frac{d{\cal B}}{dw^2}=\tau_{B_c}\frac{|\vec{p}_\pi|
|\vec{p}_D | }{32\pi^3m^3_{B_c}}|{\cal A}|^2,
\label{expr-br}
\end{eqnarray}
where $\tau_{B_c}$ is the mean lifetime of $B_c$ meson, $|\vec{p}_\pi|$ and $|\vec{p}_D|$ denote the magnitudes of
the $\pi$ and $D$ momenta in the center-of-mass frame of the pion pair,
\begin{eqnarray}
|\vec{p}_\pi|&=&\frac12\sqrt{w^2-4m^2_{\pi}}, \non
|\vec{p}_D|&=&\frac{1}{2} \sqrt{[(m^2_{B_c}-m^2_D)^2-2(m^2_{B_c}+m^2_D)w^2+w^4]/w^2}.
\end{eqnarray}

\begin{table}[tbp]
\begin{center}
\caption{The PQCD predictions for the $CP$ averaged branching ratios and the direct $CP$
asymmetries of the $B_c \to  D_{(s)} [\rho,\rho',\rho'' \to ] \pi \pi$ decays.}
\label{tab1}
\begin{tabular}{l c c} \hline\hline     %% ^{+}_{-}
\     ~~~~Mode       &    ~~      &   Results      \\  \hline
  $B^+_{c}\to D^0[\rho^+\to] \pi^+\pi^0$\;     &$~~{\mathcal B}~(10^{-5})~~$
      &\; $1.64^{+0.33}_{-0.15}(\omega_B)^{+0.21}_{-0.03}(a^0_2)^{+0.05}_{-0.16}(a^s_2)
      ^{+0.03}_{-0.09}(a^t_2)^{+0.01}_{-0.01}(C_{D_{(s)}})$\;     \\
               &  ${\mathcal A}_{CP}$
      &\; $0.20^{+0.07}_{-0.02}(\omega_B)^{+0.03}_{-0.00}(a^0_2)^{+0.07}_{-0.01}(a^s_2)
      ^{+0.14}_{-0.08}(a^t_2)^{+0.03}_{-0.03}(C_{D_{(s)}})$\;     \\
  $B^+_{c}\to D^+[\rho^0\to ]\pi^+\pi^-$\;     &$~~{\mathcal B}~(10^{-7})~~$
      &\; $6.61^{+0.96}_{-0.59}(\omega_B)^{+0.39}_{-0.21}(a^0_2)^{+0.09}_{-0.06}(a^s_2)
      ^{+1.23}_{-0.47}(a^t_2)^{+0.31}_{-0.60}(C_{D_{(s)}})$\;     \\
              &  ${\mathcal A}_{CP}$
      &\; $-0.33^{+0.08}_{-0.03}(\omega_B)^{+0.06}_{-0.04}(a^0_2)^{+0.00}_{-0.01}(a^s_2)
      ^{+0.20}_{-0.04}(a^t_2)^{+0.06}_{-0.03}(C_{D_{(s)}})$\;     \\
  $B^+_{c}\to D_s^+[\rho^0\to] \pi^+\pi^-$\; &$~~{\mathcal B}~(10^{-7})~~$
      &\; $2.63^{+0.33}_{-0.32}(\omega_B)^{+0.05}_{-0.08}(a^0_2)^{+0.00}_{-0.00}(a^s_2)
      ^{+0.00}_{-0.00}(a^t_2)^{+0.10}_{-0.10}(C_{D_{(s)}})$\;     \\
              &  ${\mathcal A}_{CP}$
      &\; $0.42^{+0.03}_{-0.00}(\omega_B)^{+0.02}_{-0.00}(a^0_2)^{+0.00}_{-0.00}(a^s_2)
      ^{+0.00}_{-0.00}(a^t_2)^{+0.04}_{-0.02}(C_{D_{(s)}})$\;     \\\hline
  $B^+_{c}\to D^0[\rho'^+\to ] \pi^+\pi^0$\;     &$~~{\mathcal B}~(10^{-6})~~$
      &\; $1.36^{+0.30}_{-0.09}(\omega_B)^{+0.05}_{-0.04}(a^0_2)^{+0.19}_{-0.14}(a^s_2)
      ^{+0.00}_{-0.02}(a^t_2)^{+0.03}_{-0.12}(C_{D_{(s)}})$\;     \\
               &  ${\mathcal A}_{CP}$
      &\; $0.12^{+0.00}_{-0.01}(\omega_B)^{+0.05}_{-0.02}(a^0_2)^{+0.05}_{-0.03}(a^s_2)
      ^{+0.03}_{-0.00}(a^t_2)^{+0.09}_{-0.02}(C_{D_{(s)}})$\;     \\
  $B^+_{c}\to D^+[\rho'^0\to]\pi^+\pi^-$\;     &$~~{\mathcal B}~(10^{-7})~~$
      &\; $1.17^{+0.13}_{-0.09}(\omega_B)^{+0.05}_{-0.02}(a^0_2)^{+0.03}_{-0.01}(a^s_2)
      ^{+0.15}_{-0.08}(a^t_2)^{+0.06}_{-0.03}(C_{D_{(s)}})$\;     \\
              &  ${\mathcal A}_{CP}$
      &\; $-0.28^{+0.09}_{-0.06}(\omega_B)^{+0.07}_{-0.03}(a^0_2)^{+0.02}_{-0.00}(a^s_2)
      ^{+0.09}_{-0.11}(a^t_2)^{+0.04}_{-0.03}(C_{D_{(s)}})$\;     \\
  $B^+_{c}\to D_s^+[\rho'^0\to ]\pi^+\pi^-$\; &$~~{\mathcal B}~(10^{-8})~~$
      &\; $1.92^{+0.28}_{-0.21}(\omega_B)^{+0.05}_{-0.07}(a^0_2)^{+0.00}_{-0.00}(a^s_2)
      ^{+0.00}_{-0.00}(a^t_2)^{+0.07}_{-0.07}(C_{D_{(s)}})$\;     \\
              &  ${\mathcal A}_{CP}$
      &\; $0.37^{+0.00}_{-0.01}(\omega_B)^{+0.02}_{-0.03}(a^0_2)^{+0.00}_{-0.00}(a^s_2)
      ^{+0.00}_{-0.00}(a^t_2)^{+0.03}_{-0.06}(C_{D_{(s)}})$\;     \\
            \hline
  $B^+_{c}\to D^0[\rho''^+\to]  \pi^+\pi^0$\;     &$~~{\mathcal B}~(10^{-7})~~$
      &\; $6.30^{+1.46}_{-0.36}(\omega_B)^{+0.10}_{-0.09}(a^0_2)^{+0.34}_{-0.38}(a^s_2)
      ^{+0.35}_{-0.02}(a^t_2)^{+0.33}_{-0.31}(C_{D_{(s)}})$\;     \\
               &  ${\mathcal A}_{CP}$
      &\; $0.06^{+0.02}_{-0.08}(\omega_B)^{+0.02}_{-0.07}(a^0_2)^{+0.06}_{-0.09}(a^s_2)
      ^{+0.04}_{-0.04}(a^t_2)^{+0.05}_{-0.01}(C_{D_{(s)}})$\;     \\
  $B^+_{c}\to D^+[\rho''^0\to ]\pi^+\pi^-$\;     &$~~{\mathcal B}~(10^{-8})~~$
      &\; $6.01^{+0.86}_{-0.57}(\omega_B)^{+0.26}_{-0.07}(a^0_2)^{+0.03}_{-0.05}(a^s_2)
      ^{+0.66}_{-0.47}(a^t_2)^{+0.33}_{-0.22}(C_{D_{(s)}})$\;     \\
              &  ${\mathcal A}_{CP}$
      &\; $-0.21^{+0.05}_{-0.03}(\omega_B)^{+0.07}_{-0.01}(a^0_2)^{+0.01}_{-0.02}(a^s_2)
      ^{+0.19}_{-0.12}(a^t_2)^{+0.07}_{-0.03}(C_{D_{(s)}})$\;     \\
  $B^+_{c}\to D_s^+[\rho''^0\to ] \pi^+\pi^-$\; &$~~{\mathcal B}~(10^{-9})~~$
      &\; $9.29^{+1.25}_{-1.17}(\omega_B)^{+0.27}_{-0.32}(a^0_2)^{+0.00}_{-0.00}(a^s_2)
      ^{+0.00}_{-0.00}(a^t_2)^{+0.14}_{-0.34}(C_{D_{(s)}})$\;     \\
              &  ${\mathcal A}_{CP}$
      &\; $0.30^{+0.01}_{-0.02}(\omega_B)^{+0.02}_{-0.02}(a^0_2)^{+0.00}_{-0.00}(a^s_2)
      ^{+0.00}_{-0.00}(a^t_2)^{+0.05}_{-0.00}(C_{D_{(s)}})$\;     \\
               \hline   \hline
\end{tabular}
\end{center}
\end{table}

Based on the decay amplitudes as given in Eqs.~(\ref{eq:a01}-\ref{eq:a1}) and the differential decay rate in Eq.~(\ref{expr-br}), we obtain
the PQCD predictions for the $CP$-averaged branching ratios (${\cal B}$) and the direct $CP$-violating
asymmetries (${\cal A}_{\rm CP}$) of the $B_c \to  D_{(s)} [ \rho, \rho',\rho'' \to ] \pi \pi$ decays, and list
the numerical results in Table~\ref{tab1}.
The first error of these PQCD predictions comes from $\omega_B=(0.60 \pm 0.05)$ {\rm GeV}
for $B_c$ meson, the following three errors are from the Gegenbauer coefficients in the two-pion distribution
amplitudes: $a^0_{2}=0.30 \pm 0.05$, $a^s_{2}=0.70 \pm 0.20$, $a^t_{2}=-0.40 \pm 0.10$ and the last error is
from $C_D=0.5 \pm 0.1$~( $C_{D_s}= 0.4\pm 0.1 $ ) in $D$ ($D_s$) meson wave function.
The total theoretical error is about $10\%$ to $30\%$ of  the central values.

\begin{figure}[tbp]
\vspace{-0.5 cm}
\centerline{\epsfxsize=8cm \epsffile{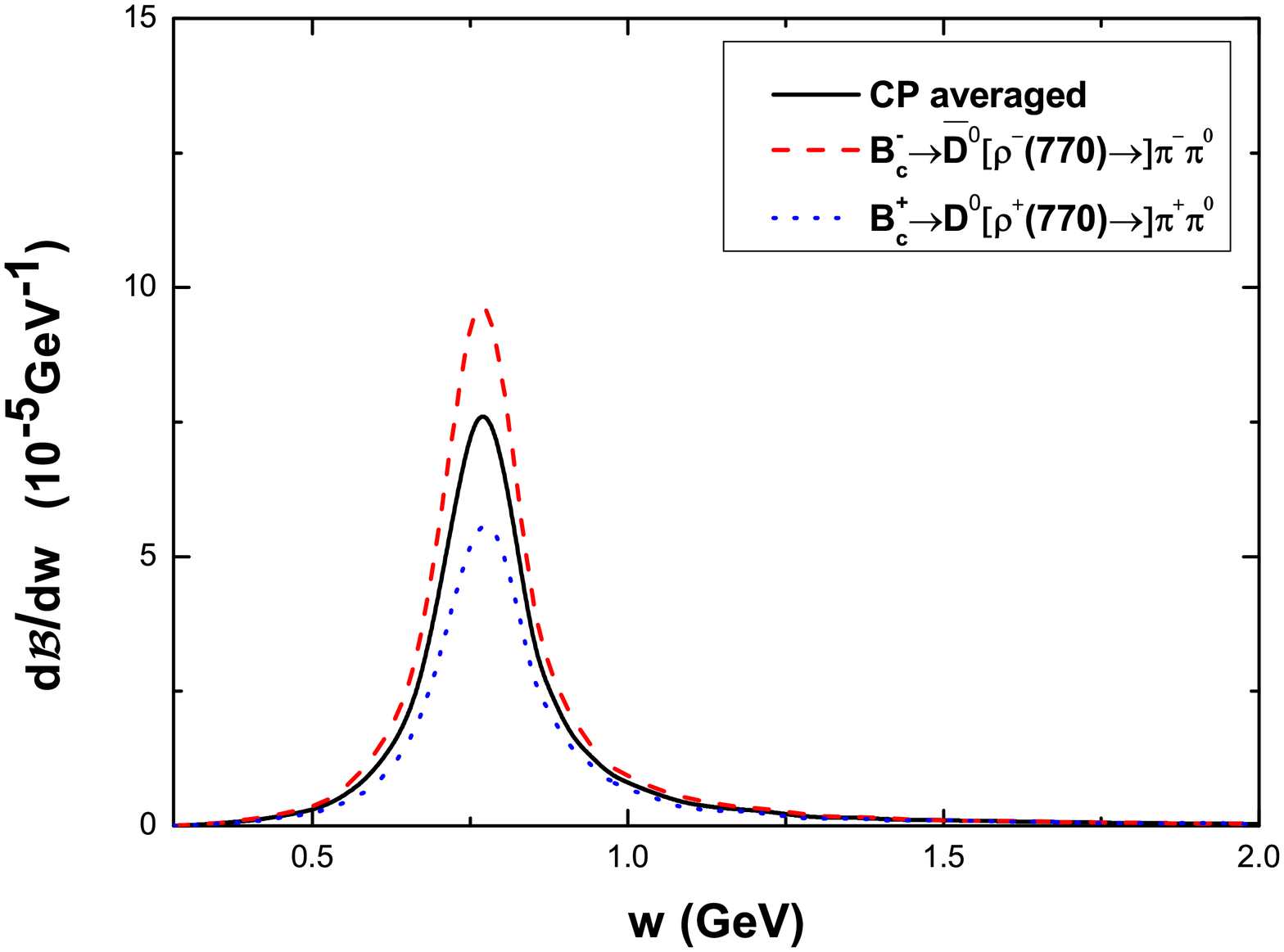}
            \epsfxsize=8cm \epsffile{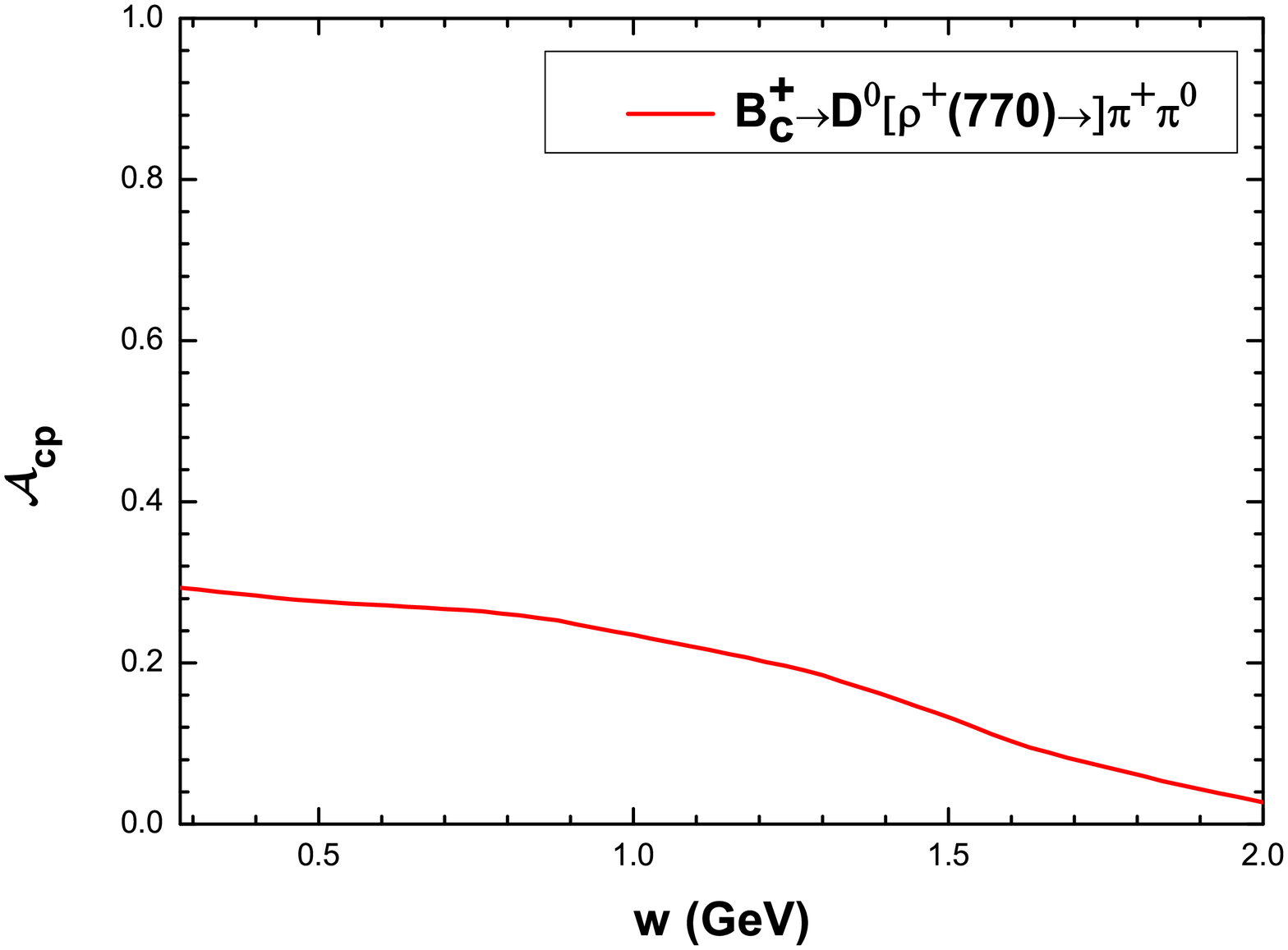}}
\vspace{-0.2cm}
  {\scriptsize\bf (a)\hspace{7.5cm}(b)}
\caption{ The PQCD predictions for $d{\cal B}/dw$ (a) and for  ${\cal A }_{CP}$ (b) for the considered
$B_c^\pm\to D^0/\bar{D^0}[\rho^\pm\to ] \pi^\pm \pi^0$ decays.}
\label{fig:fig2}
\end{figure}

\begin{figure}[tbp]
\vspace{-0.5 cm}
\centerline{\epsfxsize=8cm \epsffile{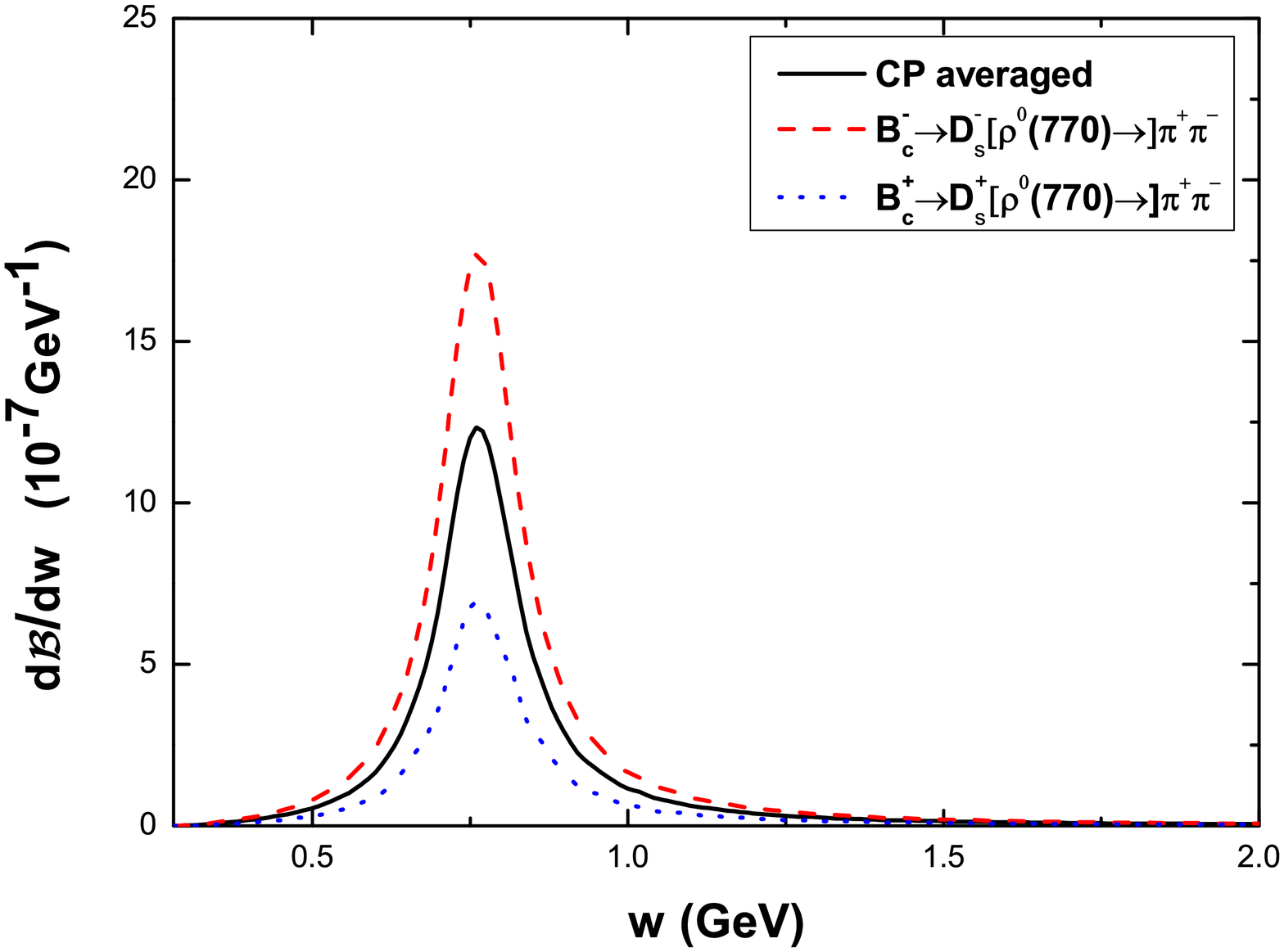}
            \epsfxsize=8cm \epsffile{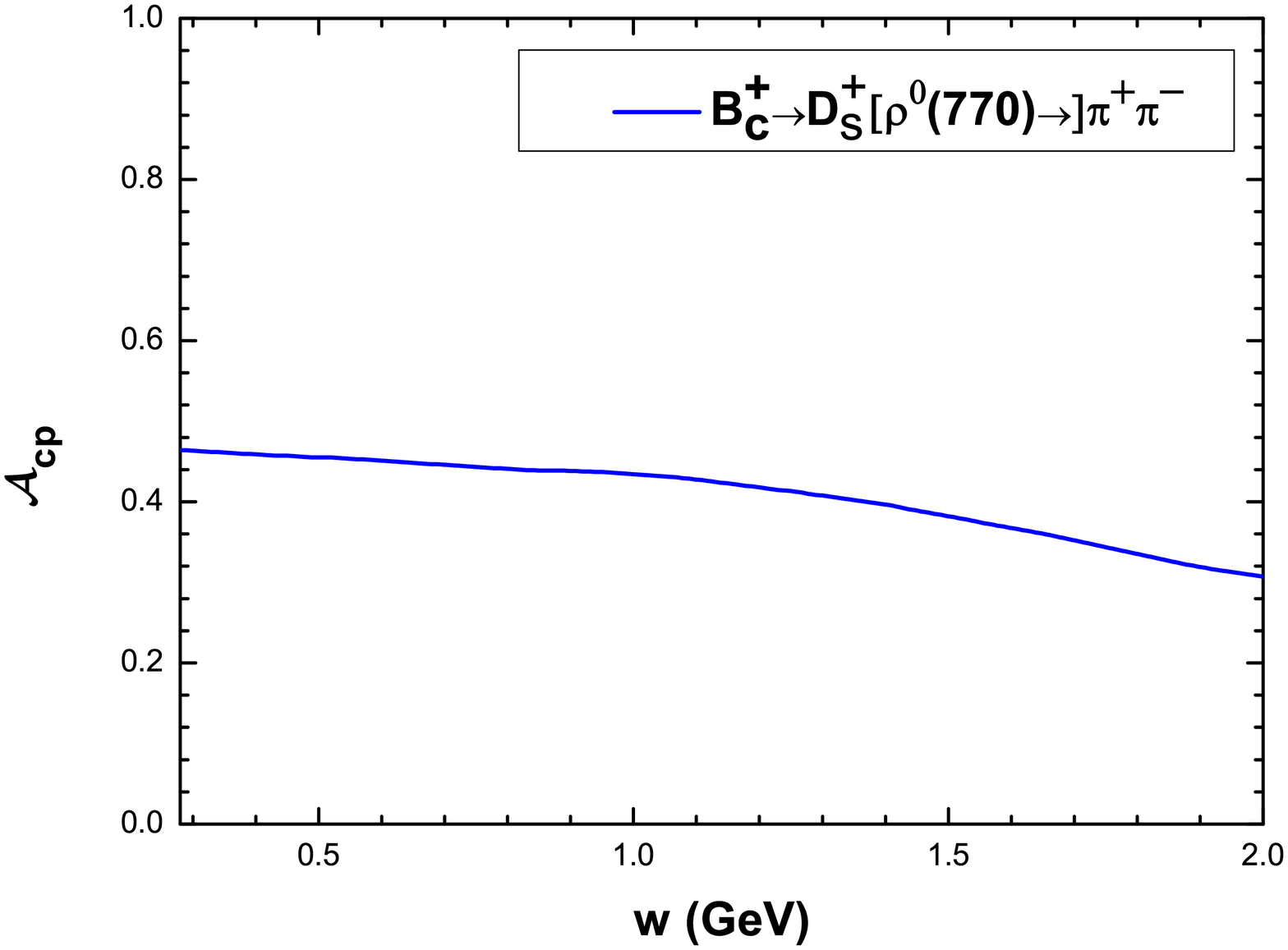}}
\vspace{-0.2cm}
  {\scriptsize\bf (a)\hspace{7.5cm}(b)}
\caption{ The same as Fig.~\ref{fig:fig2} but for $B_c^\pm\to D_s^\pm [\rho^0\to ] \pi^+\pi^-$ decays.}
\label{fig:fig3}
\end{figure}

In Fig.~\ref{fig:fig2}, we show the PQCD predictions for the differential decay rate $d{\cal B}/dw$ (Fig.~\ref{fig:fig2}(a) )
and for the $CP$-violating asymmetry ${\cal A }_{CP}$ (Fig.~\ref{fig:fig2}(b) ) for the considered
$B_c^+\to D^0[\rho^+ \to ] \pi^+ \pi^0$ decay and its charged-conjugation $B_c^-\to \bar{D}^0[\rho^- \to ] \pi^- \pi^0$
decay.  In Fig.~\ref{fig:fig3}, we show the same kinds of PQCD predictions
as in Fig.~\ref{fig:fig2} but for the $B_c^+\to D_s^+[\rho^0 \to ] \pi^+ \pi^-$ decay and its
charged-conjugation $B_c^-\to D_s^-[\rho^0 \to ] \pi^+ \pi^-$ decay.

%%========================================

%%=========================================

From the Figs.~(\ref{fig:fig2},\ref{fig:fig3}) and the numerical results as listed in Table ~\ref{tab1},
we have the following observations:
\begin{itemize}
\item[(1)]
For the considered quasi-two-body decays, the PQCD predictions  are
in the order of $10^{-9}$ to $ 10^{-5}$ for the $CP$-averaged branching ratios,  and around $(10-40)\%$
in size for the direct $CP$ violations.
The $B_c^+ \to  D^0 [\rho^+ \to ] \pi^+ \pi^0$ decay has the largest branching ratio, $\sim 1.64\times 10^{-5}$,
and to be measured in LHCb experiment.

\item[(2)]
Among the three decays involving $\rho(770)$ meson, there is a large hierarchy between their decay rates:
\beq
R_{1a} &=& \frac{ {\cal B}(B_c^+ \to  D^+ [\rho^0 \to ]\pi^+\pi^-)}{{\cal B}(B_c^+ \to  D^0 [\rho^+ \to ] \pi^+ \pi^0)}
\approx 4 \times 10^{-2}, \label{eq:r1a} \\
R_{1b} &=& \frac{ {\cal B}(B_c^+ \to  D_s^+ [\rho^0 \to ]\pi^+\pi^-)}{{\cal B}(B_c^+ \to  D^0 [\rho^+ \to ] \pi^+ \pi^0)}
\approx 1.6\times 10^{-2}, \label{eq:r1b} \\
R_{1c} &=& \frac{ {\cal B}(B_c^+ \to  D_s^+ [\rho^0 \to ]\pi^+\pi^-)}{{\cal B}(B_c^+ \to  D^+ [\rho^0 \to ] \pi^+ \pi^-)}
\approx 0.40, \label{eq:r1c}
\eeq
For the special $B^+_{c}\to D^0[\rho^+\to ] \pi^+\pi^0$ decay, the factorizable emission diagram ( i.e. the
term proportional to $a_1 F_{eD}^{LL}$ of the decay amplitude in Eq.~(\ref{eq:a01})) provide the dominant contribution.
For $B_c^+ \to  D^+ [\rho^0 \to ]\pi^+\pi^-$ decay, however, the dominant contribution comes from the
term proportional to $a_2 F_{eD}^{LL}$ of the decay amplitude in Eq.~(\ref{eq:a02}).
The small ratio $R_{1a} \approx 0.04$ can be understood basically by the strong suppression due to the ratio
$|a_2/a_1|^2 \sim 0.04$.
For $B_c^+ \to  D_s^+ [\rho^0 \to ]\pi^+\pi^-$ decay, besides the strong suppression due to $|a_2/a_1|^2$,
a new suppression factor $|V_{us}/V_{ud}|^2\sim \lambda^2 $ may also be responsible
for the ratio $R_{1c}$.

\item[(3)]
For the decay modes with the same pion pair final states  but involving the different intermediate
resonant state $\rho, \rho'$  or $\rho''$, there exists the second hierarchy between the PQCD
predictions for their decay rates.
Taking  $B_c^+  \to  D^0 [ \rho, \rho',\rho'' \to] \pi^+ \pi^0$ decays as a example, we can define the new ratios
$R_{2a,2b,2c}$:
\beq
R_{2a} &=& \frac{ {\cal B}(B_c^+ \to  D^0[ \rho'^+ \to]\pi^+ \pi^0)}{{\cal B}( B_c^+ \to  D^0 [\rho^+ \to ] \pi^+ \pi^0) }
\approx 8.3\times 10^{-2}, \label{eq:r2a} \\
R_{2b} &=& \frac{ {\cal B}(B_{c}^+ \to D^0[ \rho''^+ \to]\pi^+ \pi^0 )}{{\cal B}( B_c^+ \to  D^0 [\rho^+ \to ] \pi^+ \pi^0 )}
\approx 3.8\times 10^{-2},  \label{eq:r2b} \\
R_{2c} &=& \frac{ {\cal B}(B_{c}^+ \to D^0[ \rho''^+ \to]\pi^+ \pi^0 )}{{\cal B}( B_c^+ \to  D^0 [\rho'^+ \to ] \pi^+ \pi^0 )}
\approx 0.46. \label{eq:r2c}
\eeq
Here the main reason for the hierarchy as shown by above ratios $R_{2a,2b}$ and $R_{2c}$ is the
difference between the pion pair form factor  $F_\pi $ for different intermediate resonance $\rho$, $\rho'$ and $\rho''$.
For other two sets of decay modes, we find the similar hierarchy.
From the three curves as shown in Fig.~\ref{fig:fig4}(a), one can see directly the large difference between
the differential decay rates $d{\cal B}/dw$ for $B_c^+  \to  D_s^+ [ \rho, \rho',\rho'' \to] \pi^+ \pi^-$ decays.

%%====================================================
\begin{figure}[tbp]
\vspace{-0.5 cm}
\centerline{\epsfxsize=8cm \epsffile{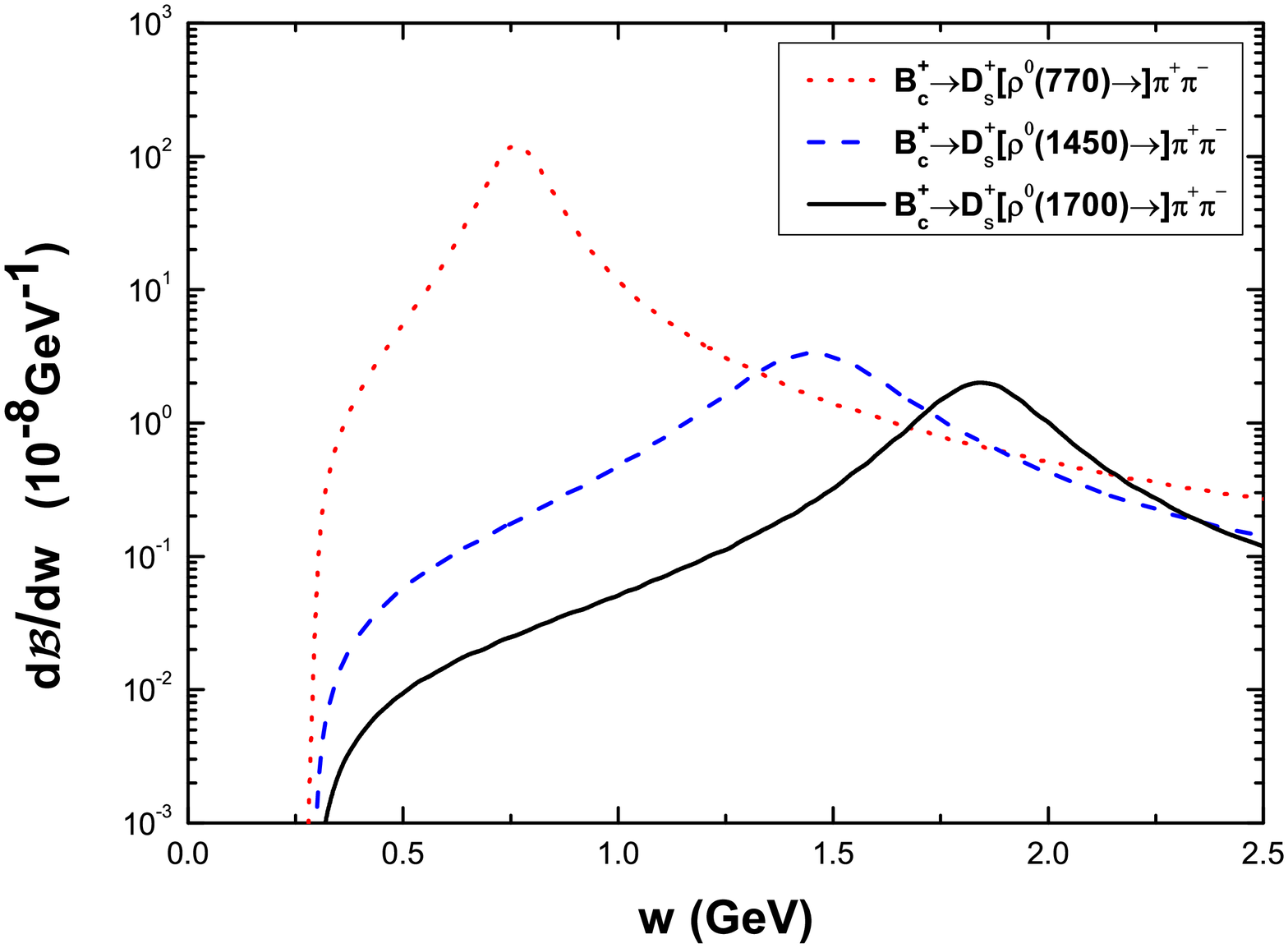}
            \epsfxsize=8cm \epsffile{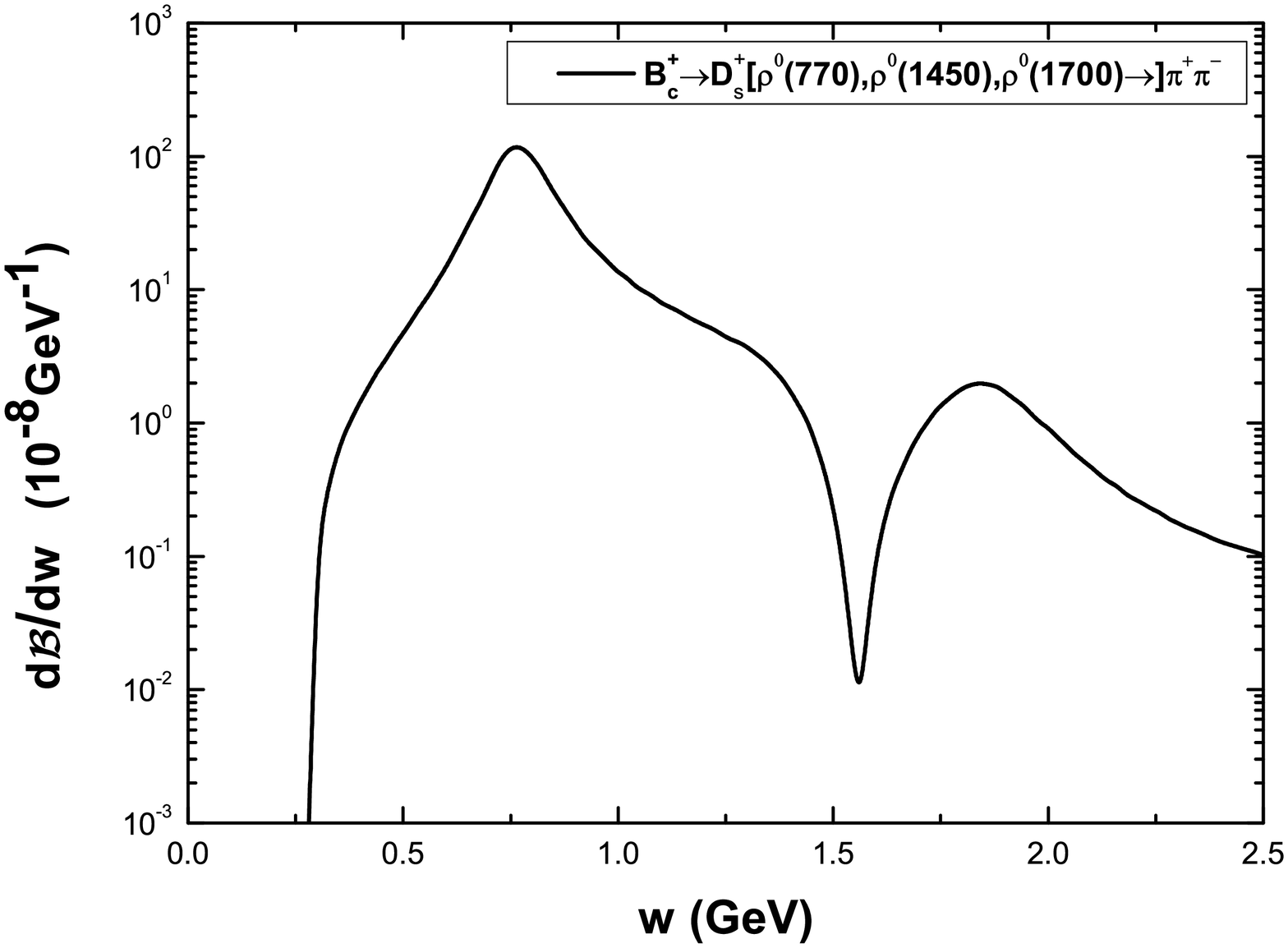}}
\vspace{-0.2cm}
  {\scriptsize\bf (a)\hspace{7.5cm}(b)}
\caption{(a) The $CP$ averaged differential decay rates for the
$B^+_{c}\to D_s^+[\rho, \rho',\rho''\to ]
\pi^+\pi^-$ decays;
and (b) The total differential decay rate when the interference
effects between $\rho$, $\rho'$ and $\rho''$ also be included.} \label{fig:fig4}
\end{figure}
%%=====================================================

\item[(4)]
By using the following well-known relation of the decay rates between the quasi-two-body and the corresponding
two-body decay modes,
\begin{eqnarray}
{\cal B}( B_c \to  D_{(s)} [ \rho( \rho', \rho'')\to ] \pi \pi  ) =
{\cal B}( B_c \to  D_{(s)} \rho (\rho', \rho'') ) \cdot {\cal B}(\rho (\rho', \rho'') \to \pi \pi),
\label{eq:def1}
\end{eqnarray}
one can extract the theoretical predictions for ${\cal B}( B_c \to  D_{(s)} \rho (\rho', \rho'') )$
from those for relevant quasi-two-body decays, if the decay rates ${\cal B}(\rho (\rho', \rho'') \to \pi \pi)$
can be determined by employing other theoretical methods or measured directly by experiments.

For the decays involving $\rho$ meson, for example, since ${\cal B}(\rho\to \pi\pi) \approx 100\%$, we therefore do
expect a similar branching ratio for the two-body $B_c \to  D_{(s)} \rho $ decay and the
corresponding quasi-two-body one.
In order to examine this general expectation, we do the calculations for ${\cal B}(B_c \to  D_{(s)} \rho) $
directly by employing the PQCD approach in the same way as Ref.~\cite{prd86-074008}. We used the
same formulae as in Ref.~\cite{prd86-074008}, but with the updated input parameters and the new  wave functions.
In the second and third column of Table \ref{tab2}, we list our PQCD predictions obtained in the framework
of the ``Quasi-two-body'' and two-body decay.
In the forth, fifth and sixth columns of the Table \ref{tab2}, as a comparison,  we also show the relevant PQCD
predictions as given  in Ref.~\cite{prd86-074008}, and the theoretical predictions obtained by using the
relativistic constituent quark model (RCQM) \cite{prd56-4133}, or by using the light-front quark model (LFQM)
\cite{prd80-114003}.

\begin{table}[tbp]
\begin{center}
\caption{In the framework of the quasi-two-body or two-body decays, we list the PQCD predictions for the $CP$ averaged
branching ratios ${\cal B}(B_c \to  D_{(s)} [\rho \to ] \pi \pi)$ decays.
As a comparison, we also list the theoretical predictions as given in  Refs.~\cite{prd86-074008,prd56-4133,prd80-114003}.}
\label{tab2}
\begin{tabular}{l| ll| c c c} \hline\hline     %% ^{+}_{-}
Decays       &  ~~Quasi-two-body&~~Two-body& PQCD~\cite{prd86-074008}~&~RCQM~\cite{prd56-4133}~&~LFQM~\cite{prd80-114003}  \\  \hline
${\cal B}(B^+_{c}\to D^0[\rho^+\to ] \pi^+\pi^0)$\;$(10^{-5})$  &\; $1.64^{+0.40}_{-0.41}$\;  &\; $1.59^{+0.18}_{-0.17}$\; & \; $0.662$\; &\; $0.60$\; &\; $0.13$\;\\
${\cal B}(B^+_{c}\to D^+[\rho^0\to ] \pi^+\pi^-)$\;$(10^{-7})$  &\; $6.61^{+1.64}_{-0.99}$\; &\; $6.28^{+1.17}_{-0.48}$\; & \; $1.4$\; &\; $3.9$\;&\; $0.2$\;  \\
${\cal B}(B^+_{c}\to D_s^+[\rho^0\to ]\pi^+\pi^-)$\;$(10^{-7})$ &\; $2.63^{+0.35}_{-0.35}$\;   &\; $2.62^{+0.34}_{-0.32}$\; &\; $0.95$\; & \; $-$\;&\; $0.02$\;\\
\hline   \hline
\end{tabular}
\end{center}
\end{table}

From Table \ref{tab2}, one can see that the PQCD predictions as listed in the column two and three
agree very well with each other. This is a new confirmation for the self-consistency between the
quasi-two-body and two-body framework of the PQCD approach for the considered $B_c$ meson decays.
Although we used the same two-body framework and the decay amplitudes as in Ref.~\cite{prd86-074008},
but one can see that the PQCD predictions obtained in this work (column three) are much larger
than those (column four) as given in Ref.~\cite{prd86-074008}, since we here used different
distribution amplitude $\phi_{B_c}(x,b)$, different  wave function $\phi_{D_{(s)}}(x,b)$  for $D_{(s)}$ meson
and the updated Gegenbauer moments, masses and decay constants as well.
In Ref.~\cite{prd86-074008}, we set $\phi_{B_c}(x,b)=\delta(x-\frac{m_c}{m_{B_c}})$.
In this paper, however, we take $\phi_{B_c}(x,b)=\delta(x-\frac{m_c}{m_{B_c}})\cdot \exp[-\omega_{B_c}^2 b^2/2]$
as given in Eq.~(7). The wave function $\phi_D(x,b)=N_D[x(1-x)]^2$ $\cdot \exp\left (-\frac{x^2 m_D^2}{2\omega_D^2}
-\frac{\omega_D^2 b^2}{2} \right ) $ used in Ref.~\cite{prd86-074008} also be very different
from the one as given in Eq.~[\ref{eq:phidxb}] of this paper.
By direct examination, we find that the dominant changes of the PQCD predictions are induced
by the difference between the wave function  $\phi_{D_{(s)}}(x,b)$ used here and the one used in Ref.~\cite{prd86-074008}.
More studies for the structure of the heavy mesons, such as $B_c$, $D$ and
$D_s$ are clearly required. Precise experimental measurements for more $B_c$ meson decays
can also test our predictions and help us to improve the theoretical framework itself.

\item[(5)]
Due to the lack of the distribution amplitudes for $\rho'$ and $\rho''$, we can not
calculate the branching ratios of the two-body decays $B_{c} \to  D\rho'$ and $ B_{c} \to D\rho''$  by using
the traditional way in the PQCD approach.
In the framework of the quasi-two-body decays, fortunately, we can extract the PQCD predictions for
the branching ratios of  the two-body decays $B_{c} \to  D \rho' $ and $B_{c} \to  D \rho''$ from the
PQCD predictions for the branching ratios of the quasi-two-body decays $B_{c} \to  D [\rho', \rho'' \to ] \pi \pi$
if we take previously determined decay rates ${\cal B}(\rho^\prime\to\pi\pi)=10.04^{+5.23}_{-2.61}\%$
and ${\cal B}(\rho^{\prime\prime} \to \pi\pi)=8.11^{+2.22}_{-1.47}\%$ ~\cite{plb763-29,1704.07566} as input.
Based on the relation as given in Eq.~(\ref{eq:def1}) and the numerical results as listed in Table
\ref{tab1}, we can then extract the PQCD predictions for the following  two-body $B_c$ meson decays:
\begin{eqnarray}
\label{br-2body-0p}
{\mathcal B}(B^+_{c}\to D^0\rho'^+)&=& 1.36^{+0.36}_{-0.21}\times 10^{-5} \;,\non
{\mathcal B}(B^+_{c}\to D^+\rho'^0)&=& 1.17^{+0.22}_{-0.13}\times 10^{-6}\;,\non
{\mathcal B}(B^+_{c}\to D_s^+\rho'^0)&=& 1.91^{+0.29}_{-0.23}\times 10^{-7}\;, \label{br-21}
\\
{\mathcal B}( B^+_{c}\to D^0\rho''^+)&=& 7.77^{+1.94}_{-0.76}\times 10^{-6} \;,\non
{\mathcal B}(B^+_{c}\to D^+\rho''^0)&=& 7.40^{+1.43}_{-0.96}\times 10^{-7}\;,\non
{\mathcal B}(B^+_{c}\to D_s^+\rho''^0)&=& 1.15^{+0.15}_{-0.14}\times 10^{-7}\;,
\label{br-22}
\end{eqnarray}
where the individual errors have been added in quadrature. These PQCD predictions will be tested
at the future LHCb experiments.

\item[(6)]
In Fig.~\ref{fig:fig4}(b), we show the total differential decay
rate after the inclusion of the contributions from all three resonant states $\rho$, $\rho'$ and $\rho''$.
From the magnitude and the shape of the curve as illustrated in \ref{fig:fig4}(b),
one can see clearly the strong destructive interference near $1.6~{\rm GeV}$: a clear dip at
$w \approx 1.6~{\rm GeV}$, similar with the one as shown in Fig.~45 of the
Ref.~\cite{prd86-032013}, where the pion form factor-squared  $|F_\pi|^2$
measured by ${\it BABAR}$ are illustrated as a function of the invariant mass of the pion pair in the range
from $0.3$ to $3$ GeV.
In our work, the same dip is induced by  the strong destructive interference
between $\rho'$ and $\rho''$, as shown in Fig.~\ref{fig:fig4}(b).
Numerically, the PQCD predictions for the individual decay rate of $\rho'$ and $\rho''$  and the interference
term between them are the following:
\begin{eqnarray}
{\cal B}(B^+_c\to D_s^+[ \rho'\to ] \pi^+\pi^-) &\approx&1.92\times 10^{-8}\;, \non
{\cal B}(B^+_c\to D_s^+[ \rho''\to ] \pi^+\pi^-) &\approx& 9.29\times 10^{-9}\;, \non
{\rm interf. \ \  term}|_{\rho'-\rho''} &\approx &  -1.75\times 10^{-8}\;.
\label{eq:inter}
\end{eqnarray}
It is easy to see that the interference term is indeed large and negative when compared with other two individual
contributions.
\end{itemize}

\section{Summary}\label{sec:4}

In this paper, we studied the quasi-two-body $B_c \to  D_{(s)} [\rho,\rho', \rho'' \to ] \pi \pi$ decays
in PQCD factorization approach. The two-pion distribution amplitudes have been applied to include the
final-state interactions between the pion pair. The contributions from the $ \rho$, $\rho'$ and $\rho''$
intermediate resonant states were estimated by introducing the time-like form factor $F_\pi$ involved in
the $P$-wave two-pion distribution amplitudes. The PQCD predictions for the $CP$-averaged branching
ratios and direct $CP$-violating asymmetries of  the considered quasi-two-body decays are obtained
and listed in  Table \ref{tab1} and \ref{tab2}.
Based on the relation as given in Eq.~(\ref{eq:def1}),  we extract the theoretical predictions for the
branching ratios of the two-body decays $B_c \to  D_{(s)} X $ with $X=(\rho,\rho',\rho'')$
from those PQCD predictions for ${\cal B}(B_c \to  D_{(s)} [\rho,\rho', \rho'' \to ] \pi \pi)$ and those previously
determined decay rates ${\cal B}(\rho^\prime\to\pi\pi)$ and ${\cal B}(\rho^{\prime\prime} \to \pi\pi)$.

From the analytical analysis and numerical calculations, we found the following points:
\begin{itemize}
\item[(1)]
The PQCD predictions for the branching ratios of the quasi-two-body $B_c \to  D_{(s)}  [\rho,\rho', \rho''
\to ] \pi \pi$ decays are in the order of $10^{-9}$ to $ 10^{-5}$, the direct $CP$ violations are
around $(10-40)\%$ in magnitude. The decay mode $B_c^+ \to  D^0 [\rho^+ \to ] \pi^+ \pi^0$ has a large branching ratio,
$\sim 1.64\times 10^{-5}$, and could be measured in the future LHCb experiment.

\item[(2)]
The two sets of the large hierarchy $R_{1a,1b,1c}$ for the ratios between the branching ratios
${\cal B}(B_{c} \to  D_{(s)} [\rho \to ] \pi \pi)$
and $R_{2a,2b,2c}$ among the branching ratios ${\cal B}(B_c^+  \to  D^0 [ \rho, \rho',\rho'' \to] \pi^+ \pi^0)$
are defined and can be understood in the PQCD factorization approach.
The self-consistency between the quasi-two-body and two-body framework for
$B_{c} \to  D_{(s)} [\rho  \to ] \pi \pi$ and $B_{c} \to  D_{(s)} \rho$ decays are confirmed by our
numerical  results.

\item[(3)]
Taking previously determined decay rates ${\cal B}(\rho^\prime\to\pi\pi)\approx 10\%$
and ${\cal B}(\rho^{\prime\prime} \to \pi\pi)\approx 8.1\%$ as input,
we extract the theoretical predictions for branching ratios  ${\cal B}(B_{c} \to  D \rho') $
and ${\cal B}(B_{c} \to  D \rho'')$ from the PQCD predictions for the branching ratios of
the quasi-two-body decays $B_{c} \to  D [\rho'\to ] \pi \pi$ and $B_{c} \to  D [\rho'' \to ] \pi \pi$.

\end{itemize}

\begin{acknowledgments}

Many thanks to Hsiang-nan Li and Wen-Fei Wang for valuable discussions.
This work is supported by the National Natural Science Foundation of China under the
No.~11235005 and 11775117. Ai-Jun Ma and Ya Li are also supported by Postgraduate Research \& Practice Innovation  Program of  Jiangsu Province under Grant No.~KYCX17-1056 and No.~KYCX17-1057.

\end{acknowledgments}

\appendix

\section{Some relevant functions}  \label{sec:app}

The explicit expressions of the evolution factors $E_e(t)$, $E_a(t)$ and $E_n(t)$ and the threshold
resummation factor $S_t(x)$ can be found, for example, in Refs.~\cite{npb-923,wang}.
We here show the explicit expressions of the hard functions $h_i$ with $i=(a,\cdots,h)$ which are
obtained from the Fourier transform of the hard kernels:
\begin{eqnarray}
h_i(\alpha,\beta,b_1,b_2)&=&h_1(\beta,b_2)\times h_2(\alpha,b_1,b_2),\nonumber\\
h_1(\beta,b_2)&=&\left\{\begin{array}{ll}
K_0(\sqrt{\beta}b_2), & \quad  \quad \beta >0\\
K_0(i\sqrt{-\beta}b_2),& \quad  \quad \beta<0
\end{array} \right.\non
h_2(\alpha,b_1,b_2)&=&\left\{\begin{array}{ll}
\theta(b_2-b_1)I_0(\sqrt{\alpha}b_1)K_0(\sqrt{\alpha}b_2)+(b_1\leftrightarrow b_2), & \quad   \alpha >0;\\
\theta(b_2-b_1)I_0(\sqrt{-\alpha}b_1)K_0(i\sqrt{-\alpha}b_2)+(b_1\leftrightarrow b_2),& \quad   \alpha<0;
\end{array} \right.
\end{eqnarray}
where $K_0$ and $I_0$ are modified Bessel functions with
$K_0(ix)=\frac{\pi}{2}(-N_0(x)+i J_0(x))$ and $J_0$ is the Bessel function.
The hard scale $t_i$ is chosen as the maximum of the virtuality of the internal momentum transition in the hard amplitudes:
\begin{eqnarray}
t_a&=&\max \{m_B\sqrt{|\alpha_a|},m_B\sqrt{|\beta_a|}, 1/b_3, 1/b_B \},~~~~
t_b=\max \{m_B\sqrt{|\alpha_b|},m_B\sqrt{|\beta_b|}, 1/b_B, 1/b_3 \};\nonumber\\
t_c&=&\max \{m_B\sqrt{|\alpha_c|},m_B\sqrt{|\beta_c|}, 1/b_B, 1/b \},~~~~~
t_d=\max \{m_B\sqrt{|\alpha_d|},m_B\sqrt{|\beta_d|}, 1/b_B, 1/b \};\nonumber\\
t_e&=&\max \{m_B\sqrt{|\alpha_e|},m_B\sqrt{|\beta_e|}, 1/b_3, 1/b \},~~~~~~
t_f=\max \{m_B\sqrt{|\alpha_f|},m_B\sqrt{|\beta_f|}, 1/b, 1/b_3 \};\nonumber\\
t_g&=&\max \{m_B\sqrt{|\alpha_g|},m_B\sqrt{|\beta_g|}, 1/b, 1/b_B \},~~~~~
t_h=\max \{m_B\sqrt{|\alpha_h|},m_B\sqrt{|\beta_h|}, 1/b, 1/b_B \},\nonumber\\
\end{eqnarray}
where
 \begin{eqnarray}
\alpha_a&=& r_b^2+(1-r^2)[(\eta-1)x_3-\eta],~~~~~~~\beta_a= (r^2-x_B)[(1-\eta)(x_3-1)+x_B];\nonumber\\
\alpha_b&=& (r^2-x_B ) (x_B+\eta-1 ),~~~~~~~~~~~~~~~~\beta_b=\beta_a;\nonumber\\
\alpha_c&=&\beta_a,~~~~~~~~~~~~~~~~~~~~~~~~~~~~~~~~~~~~~~~~~~~~~~~~~~~\beta_c= [1-x_B-(1-r^2)z][(1-\eta)x_3+x_B-1];\nonumber\\
\alpha_d&=&\beta_a,~~~~~~~~~~~~~~~~~~~~~~~~~~~~~~~~~~~~~~~~~~~~~~~~~~~\beta_d= [(1-z)r^2-x_B+z][(1-\eta)(x_3-1)+x_B];\nonumber\\
\alpha_e&=& (1-r^2)[(\eta-1)x_3-\eta],~~~~~~~~~~~~~~~~\beta_e=(1-\eta)(r^2-1)x_3z;\nonumber\\
\alpha_f&=& r_c^2+(1-\eta)[r^2(z-1)-z],~~~~~~~~~\beta_f=\beta_e;\nonumber\\
\alpha_g&=&\beta_e, ~~~~~~~~~~~~~~~~~~~~~~~~~~~~~~~~~~~~~~~~~~~~~~~~~~~\beta_g= r_b^2-[(1-r^2)z+x_B-1][(1-\eta)x_3+x_B-1];\nonumber\\
\alpha_h&=&\beta_e,~~~~~~~~~~~~~~~~~~~~~~~~~~~~~~~~~~~~~~~~~~~~~~~~~~~\beta_h= r_c^2-[(r^2-1)z+x_B][(\eta-1)x_3+x_B].
\end{eqnarray}

%========================= reference=========================%

\end{document}